# Impact of dopant species on the interfacial trap density and mobility in amorphous In-X-Zn-O solution-processed thin-film transistors

Mohammed Benwadih,[1,a)] J. A. Chroboczek,[2] Gerard Ghibaudo,[2] Romain Coppard,[1] and Dominique Vuillaume[3]
[1]*CEA Grenoble/DRT/Liten, 17 rue des martyrs 38054, Grenoble, France*
[2]*IMEP-LAHC, MINATEC-INPG, 3 rue Parvis Louis Néel, 38016 Grenoble, France*
[3]*Institute for Electronics Microelectronics and Nanotechnology, CNRS, Avenue Poincaré, 59652, Villeneuve d'Ascq, France*

Alloying of In/Zn oxides with various X atoms stabilizes the IXZO structures but generates electron traps in the compounds, degrading the electron mobility, $\mu$. To assess whether the latter is linked to the oxygen affinity or the ionic radius, of the X element, several IXZO samples are synthesized by the sol-gel process, with a large number (14) of X elements. The IXZOs are characterized by XPS, SIMS, DRX, and UV-spectroscopy and used for fabricating thin film transistors. Channel $\mu$ and the interface defect density $N_{ST}$, extracted from the TFT electrical characteristics and low frequency noise, followed an increasing trend and the values of $\mu$ and $N_{ST}$ are linked by an exponential relation. The highest $\mu$ (8.5 cm$^2$V$^{-1}$s$^{-1}$) is obtained in In-Ga-Zn-O, and slightly lower value for Sb and Sn-doped IXZOs, with $N_{ST} \approx 2 \times 10^{12}$ eV$^{-1}$ cm$^{-2}$, close to that of the In-Zn-O reference TFT. This is explained by a higher electronegativity of Ga, Sb, and Sn than Zn and In, their ionic radius values being close to that of In and Zn. Consequently, Ga, Sb, and Sn induce weaker perturbations of In-O and Zn-O sequences in the sol-gel process, than the X elements having lower electronegativity and different ionic radius. The TFTs with X = Ca, Al, Ni and Cu exhibited the lowest $\mu$ and $N_{ST} > 10^{13}$ eV$^{-1}$cm$^{-2}$, most likely because of metallic or oxide clusters formation.

## I. INTRODUCTION

High mobility thin-film transistors (TFTs) with ZnO, In-Ga-Zn-O (IGZO), In-Zn-O (IZO), Zn-Sn-O (ZTO),[1–5] channels have attracted a considerable attention because of their capacity of being used in flat panel displays, solar cells, wide area sensors, and various flexible electronics circuitry. Oxide semiconductors have numerous advantages over conventional amorphous silicon and organic semiconductor materials.[6] They feature long-range material uniformity thanks to their strictly amorphous structure, are environmentally stable, and show electron mobility, as high as 20 cm$^2$V$^{-1}$s$^{-1}$. And, last but not least, they are transparent in visible light, which opens a vast field of applications.

Metal oxide semiconductor films are usually grown by vapor deposition techniques, such as pulsed laser material ablation, or rf-magnetron sputtering. The TFTs fabricated using such semiconducting oxides have rather satisfactory characteristics, for example, electron mobility exceeding 15 cm$^2$ V$^{-1}$ s$^{-1}$, $I_{on}/I_{off}$ ratio above 10$^7$, and stable threshold voltages, $V_{th}$, close to 0. However, these vapor deposition techniques require using expensive high-vacuum equipment and impose limits on the sample size, because of restricted volume of the deposition chambers.

In this work, oxide semiconductors films were grown by spin coating using a solution (sol-gel) technique. That method has several advantages over vacuum-assisted techniques, such as high throughput, cost-effectiveness, low prime material waste and above all, simplicity. It is also well-adapted for manufacturing materials or devices on large surfaces. For these reasons, the use of solution-processed materials for fabricating transparent metal oxides devices is becoming increasingly popular.

In spite of these advantages, sol-gel based devices often show severe bias, light, and environmental instabilities. Very recently, several research groups successfully processed ZTO, IGZO, and IZO TFTs, using sol-gel and spin coating techniques.[4] These devices, show electron mobility from 1 cm$^2$V$^{-1}$s$^{-1}$ to 20 cm$^2$V$^{-1}$s$^{-1}$, depending on the material chemical composition, doping, structure, and a post process handling.

Electron mobility in the oxides is known to depend on the chemical composition of the material. This is attributed to the specific electronic band structure of amorphous metal oxide semiconductors. Namely, the conduction band originates from the unoccupied s-orbitals of heavy post-transition metal cations. At a sufficiently high concentration of the latter, their (n − 1)d$^{10}$ns$^0$ orbitals overlap sufficiently to delocalize the ns-electrons. The delocalization occurs more easily, i.e., at a lower dopant concentration, for cations with a higher spatial extent of their ns$^0$ wave functions, which grows with the increasing principal quantum number n. In fact, it has been observed that the delocalization occurs only for cations with n ≥ 5. When all these conditions are met, the conduction band can be wide enough to assure high electron mobility.

The main reason to use indium as a constituent of the conducting IZO is that In$^{3+}$ is a heavy post-transition metal

---

[a)]Author to whom correspondence should be addressed. Electronic mail: mohammed.benwadih@cea.fr.

cation with the electronic configuration of $[Kr](4d)^{10}(5s)^0$. Note that its principal quantum number n is 5, whereas the $Zn^{2+}$ ion has an electronic configuration of $[Ar](3d)^{10}(4s)^0$, with n = 4. Therefore, the s-orbital of $In^{3+}$ extends farther than that of $Zn^{2+}$, thus the mean overlap of s-orbitals centered at neighboring cation sites becomes more pronounced, which makes the mixed-oxide film more conductive. This simple model explains the observed increase in both the on-current level and the electron mobility, as the indium concentration in the sol-gel-derived IGZO TFTs is increased.[2,6–9]

However, IZO shows structural instabilities. They can be controlled by alloying it with a fourth component, at significant concentrations. It is called a dopant, like in inorganic semiconductors, but it is added at concentration orders of magnitude higher than in the latter, reaching the level of a few atomic %. Its role is multiple; it is not only an electron donor, but it also stabilizes the material structure and increases or decreases conduction in chains of s-orbitals, as mentioned above. A correct choice of a dopant is crucial for the material properties and numerous teams have worked on this topic. The idea underlying all the research on doping of IZO is to find an element X that perturbs the least the origi-nal system structure, i.e., it retains its conductive properties, but renders it, at the same time, more stable. It has been generally agreed that Ga is the element closest to such expectations and IGZO is today's best known ternary oxide semiconductor.[10,11] However, the effect of Ga-doping of In-Zn-O on the mobility and the $I_{on}$ current level is more complex. Gallium in IZO also plays a role of a "getter," that is an oxygen-retaining agent. It is so because oxygen affinity to gallium is higher than that to indium or zinc. In other words, the Ga-O bonds are stronger than the In-O or Zn-O bonds. Because of the affinity difference, gallium retains oxygen, which results in a decrease in the oxygen vacancy concentration, entailing in turn a decrease in the charge carriers density. Thus, at a higher Ga concentration, the material resistivity increases.[12,13] Several teams have studied also the effect of aluminum or magnesium incorporation on the performance of the IZO matrix and on its morphology.[14,15] Kamiya et al.[16] studied the effects of defects and the doping on carrier mobility and showed that they are generally significant and depend on the annealing temperature and on the dopant species. They argue that the defects and also the dopants (Ga, H) affect the configuration of oxygen atoms, changing the transport properties of the material.[16]

In this work, we carried out a systematic study on In-X-Zn-O (henceforth abbreviated to IXZO) properties, for 14 dopant species, X = Ga, Sb, Sn, Mg, Be, Ag, Y, Ca, Al, Ni, Cu, Mn, Mo, and Pd. The X concentrations were kept constant and lower than 2 at. % in order to point out principal factors affecting transport properties of the materials. We studied the composition of the IXZO layers using X-ray photoelectron spectroscopy (XPS) and the distribution of elements in the film by Time-of-Flight Secondary Ion Mass Spectrometry (Tof-SIMS). The material structure was characterized by X-ray diffraction (XRD).

In our dopant list, we can identify species with various affinity to oxygen; some ions are more electronegative ($Ga^{3+}$, $Sn^{2+}$, and $Sb^{3+}$) than Indium and Zinc and others, such as ($Be^{2+}$, $Mg^{2+}$, $Ca^{2+}$, and $Y^{3+}$) ions are less electronegative. As it can be readily seen, this list includes also several transition metal ions ($Mo^{2+}$, $Mn^{2+}$, $Ni^{2+}$, $Cu^{2+}$, $Pd^{2+}$, $Ag^+$), of intermediate electronegativity, the dopants X are ranked there according to their ionic radius (IR) values, (from the smallest to the highest).[17]

The objective of this work is to determine what dopant properties are significant for the electrical performances of the related TFTs. The properties of the IXZO materials are compared with metal oxides showing the best, well-established electrical performances, i.e., IGZO and IZO.

All the studied IXZO films were fabricated maintaining the In/Zn ratio at 1/2 as such composition assures electrically stable and reproducible layer formation which results in satisfactory TFT characteristics (low $I_{off}$ and correct subthreshold swing). Kim et al.[11] have shown that at higher In concentration, the $I_{off}$ current increases sharply with a simultaneous subthreshold slope deterioration, mobility increase and a shift of $V_{th}$ to negative voltages. Banger et al.[15] studying devices with several different In/Zn ratios, all annealed at 550 °C, have shown that the hysteresis of I-V curves became more severe as the indium content in devices was increased. The effect of In and Ga incorporation was also studied on vacuum deposited films, and the same trends as those seen in the solution processed IGZO TFTs were reported, i.e., an increase in mobility, subthreshold slope degradation and I-V curves shift to negative voltages.[18]

The main objective of this work is the fabrication of the TFTs with IXZO channels and their electrical characterization in conjunction with studies of the physicochemical properties of the IXZO films. The TFTs were tested using standard transistor analyzing tools. Their static characteristics, $I_{DS}(V_{DS})$ and $I_{DS}(V_G)$ were used for determining basic typical transistor parameters, such as threshold voltage $V_{th}$, transconductance, $g_m$, subthreshold swing, SS of $I_{DS}(V_G)$, from which we further extracted channel carrier mobility, $\mu$, and the interface trap density, $N_{ST}$, in the channel. The latter was also obtained from an independent measurement of low frequency noise (LFN) spectral power density characteristics.

## II. OXIDE SEMICONDUCTOR FILMS

### A. Synthesis

We employed the sol-gel process for the oxide deposition and doping, for the reasons discussed above. It involved the use of a precursor, such as acetate or nitrate, chloride, in an appropriate solvent salt of the metal of interest. Again, additives could be introduced to either improve the solubility or promote the subsequent chemical conversion in the metal-oxygen−metal (M−O−M) network via hydrolysis and condensation reactions.

All precursor materials were purchased from Sigma-Aldrich and used without further purification. We used the same acetate-based precursor for all elements involved, for easier comparisons of properties of the synthesized ternary oxides.[19] That also facilitated the understanding of the mechanisms of incorporation of a X dopant into the IXZO matrix. More precisely, we used zinc acetate dehydrate, indium acetate, X acetate, and ethanolamine in anhydrous 2-methoxyethanol for preparing the precursor solution for the IXZO synthesis. The molar ratio of indium, X, zinc (In:X:Zn) in the precursor solution was kept constant at the 1:0.1:2 ratio. The indium concentration was maintained at 0.2 M and the concentration of zinc was fixed at 0.4 M. The dopant's concentration was kept constant at 0.02 M. The molar ratio of ethanolamine to indium and zinc was maintained at 1:8. The solution was stirred at 70 °C in air for 1 h and aged for 12 h in air prior to the synthesis.

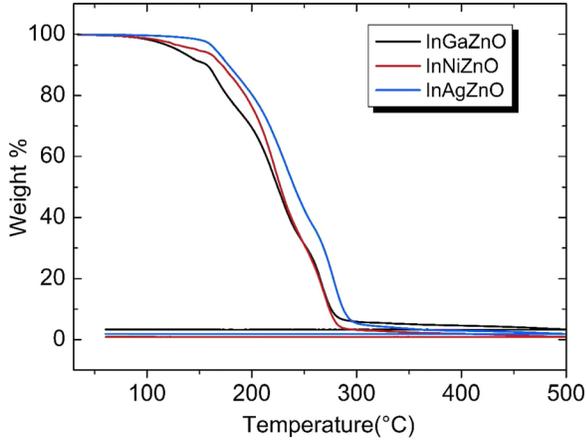

FIG. 1. TGA curves of the IGaZO, IAgZO, INiZO ink.

We carried out thermogravimetric analysis (TGA) on the metal oxide solution to check its thermal decomposition to define the optimal annealing temperature of the film (Figure 1). TGA were performed on 40–45 mg of materials by using a Pegasus Netzsch thermal analysis instrument under air with a flow rate of 10 ml/min and a heating ramp rate of 10 °C/min from 25 to 500 °C.

No perceptible weight loss was detected in the solutions at temperatures higher than 425 °C. The final weight loss is attributed to dehydroxylation of M-OH, and the removal of residual organics. TGA indicated that the thermal removal was complete for the organic compounds at around 450 °C.

The substrates were cleaned by ultrasonic acetone bath to remove the polymer protection layer, rinsed directly in deionized water, acetone, and isopropanol and dried. Just before spin-coating, the substrates were treated with oxygen plasma (power of 50 W during 60 s) to obtain a clean and residue-free surface. The precursor solution of IXZO was spin coated onto substrates at 2000 rpm for 25 s and the samples were subsequently placed onto a hot plate in air at 450 °C for 10 min in order to decompose the precursor and form the metal oxide layer. This process was repeated three times in order to obtain the desired film thickness (about 12 nm). When glass substrates were used, the solution was spin-coated at 2000 rpm for 15 s and the process repeated 5 times (thickness of about 40 nm). Then, the prepared IXZO solutions were baked at 450 °C for 1 h to form active layers in TFTs.

### B. XPS analysis

X-ray photoelectron spectroscopy (XPS) measurements were carried out to establish the chemical and local bonding configuration and structural details of the studied ternary oxides. XPS was performed on a Thermo Electron Scientific ESCALAB 250 spectrometer (the source Al Kα 1486.6 eV) at pressure of $4.5 \times 10^{-10}$ Torr (UHV), XPS spectra were obtained immediately after film fabrication in order to minimize surface contamination.

Some of the obtained XPS spectra are shown in Figure 2(a) for indium and Figure 2(b) for oxygen. Figure 2(c) shows results of a deconvolution of the oxygen 1 s spectra of IGZO into three principal components attributed to (i) the M−O−M lattice species, at 530.3 ± 0.1 eV, associated with the $O^{2-}$ ions binding with X, In, and Zn ions, without oxygen vacancies, (ii) bulk and surface metal hydroxide (M−OH) species, at 531.2 ± 0.1 eV, containing non-stoichiometric oxide species. The peak originates from the oxygen vacancies which supply free electron carriers in the ZnO-based films.[20] Finally, (iii) the weakly bound (M−OR) species such as surface adsorbed carbon species, at 532.5 ± 0.1 eV, represent the chemisorbed oxygen at the surface of IXZO.[21,22]

The band at around 530 eV is attributed to the oxygen atoms located in the vicinity of oxygen vacancies.[20] The XPS spectrum shows clearly that the films with Ga, Mg, Al, and Pd contain an appreciable concentration of oxygen vacancies. The M-O-M line can be assigned to the oxygen in the lattice and to the oxygen vacancy. The line positions for various X are the following; Ga: 530.3 eV, Be: 530.2 eV Mg: 530.1 eV, Al: 530.7 eV, and Pd: 530.1 eV.

The formation of oxygen vacancies is associated with a generation of charge carriers, according to

$$O_0^x = \frac{1}{2}O_{2(gas)} + V_0^{\circ\circ} + 2e^-. \quad (1)$$

Equation (1) states that the $O_2$ removal from the oxide sublattice ($O_0^x$) causes the appearance of a doubly charged vacancy and a release of two free electrons.

This implies that IGZO films annealed at higher temperatures should have higher vacancy concentrations and higher current charge densities.[13]

We have found that the intensity of the O1s line in the XPS spectrum of IGZO is higher than in IXZO with X = Be, Mg, Al, and Pd, as shown in Figure 2(b). That implies a higher concentration of oxygen vacancies in IGZO, hence a higher electron density and conductivity. The good performances of the TFTs with IGZO channel can be attributed to the latter. The component at 531.2 eV, attributed to M-OH bond, is relatively weak, but present for all X. all the elements were in their oxide form. The spectrum 3d lines of indium oxides (Figure 2(a)) were not affected by the presence of the dopants (Ga, Mg, Al, Pd). The In3d line is centered at 445.2 for IGZO, 445.2 eV for IMgZO 445.4 eV for IAlZO, and 444.9 eV for IPdZO. It has been previously reported that the In3d peak, centered at 444.6 eV, corresponds to the In–O bond.[22]

The spectral line energies listed above are slightly shifted towards lower energies (Figure 2(a)). These shifts arise from differences in the environment of indium in the structure and/or different metal and oxygen bonding states.

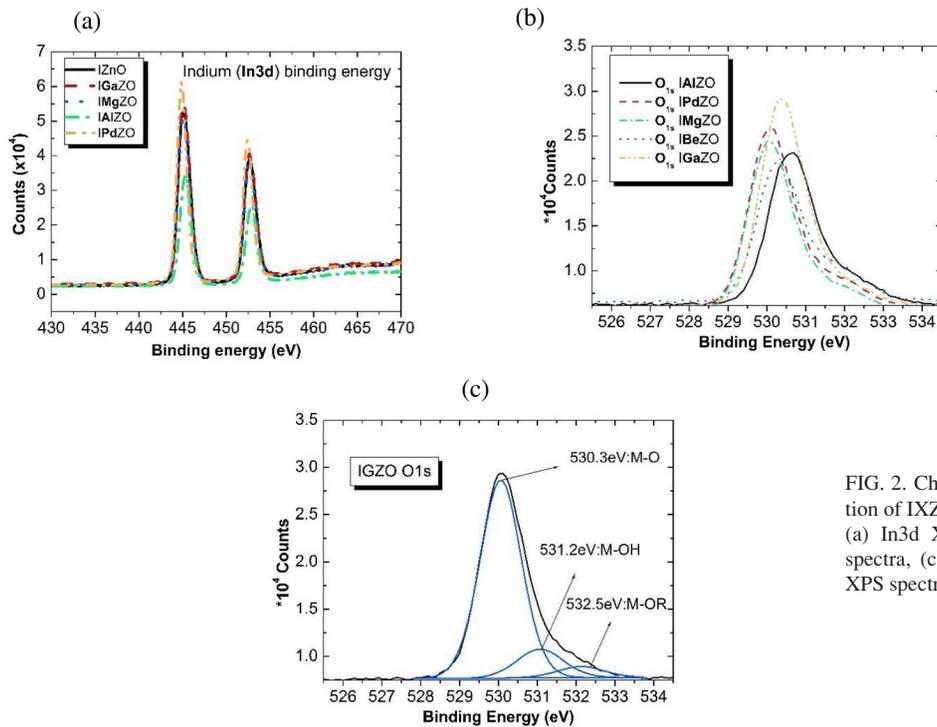

FIG. 2. Chemical and structural evolution of IXZO for (X = Ga, Mg, Al, Pd), (a) In3d XPS spectra, (b) O1s XPS spectra, (c) deconvolution of the O1s XPS spectra for IGZO.

## C. SIMS analysis

We also analyzed the in-depth distribution of X in the amorphous oxide semiconductor films. The measurements were performed (To5 instrument from IontoF Co.) in order to analyze precisely the distribution of elements in the IXZO layer on $SiO_2$ substrate. Each sample was analyzed near its center, over the area of $80 \times 80 \mu m^2$, with $O_2^+$ abrasion at 500 eV.

The SIMS analysis demonstrated (Figure 3(a)) that the distribution of dopants was homogeneous for Ga, Mg, Al, and Pd. The depth variation of all the X elements with distance, measured from the surface towards the interface is flat for all the films and the dispersion of dopant atoms in the thickness was homogeneous without signs of segregation.

The Tof-SIMS data did not reveal any significant differences in the in-depth distribution of indium (In) (Figure 3(b)) and no contamination by carbon or chlorine was detected. The thickness of the doped layers estimated by SIMS was approximately 14 nm and found uniform and close to that determined from the TEM image.

## D. XRD analysis

X-ray diffraction (XRD) measurements were performed with BRUKER D8 Advance diffractometer using Cu Kα radiation and grazing incidence was carried out with a Rigaku model ATX-G Thin-Film Diffraction Workstation using Cu Kα radiation, coupled to a multilayer mirror. The XRD patterns showed that all the IXZO films, on glass substrate were amorphous, regardless of their chemical composition (Figure 4(a)). We found that the doping at concentrations lower than 2 at. % had no effect on the crystallographic structure of the semiconductor metal-oxide IXZO.

Long time (up to 36 h) XRD measurements were also carried out at a grazing incidence angle (0.15°) for IXZO films deposited on oxidized silicon substrates. Their X-ray diffraction patterns show no crystallization peaks (Figure 4(b)). All the films were found to be essentially amorphous at a macroscopic level, regardless of their composition. This result confirms previous reports stating that small amounts of dopants do not affect the crystallinity of the structures.[21]

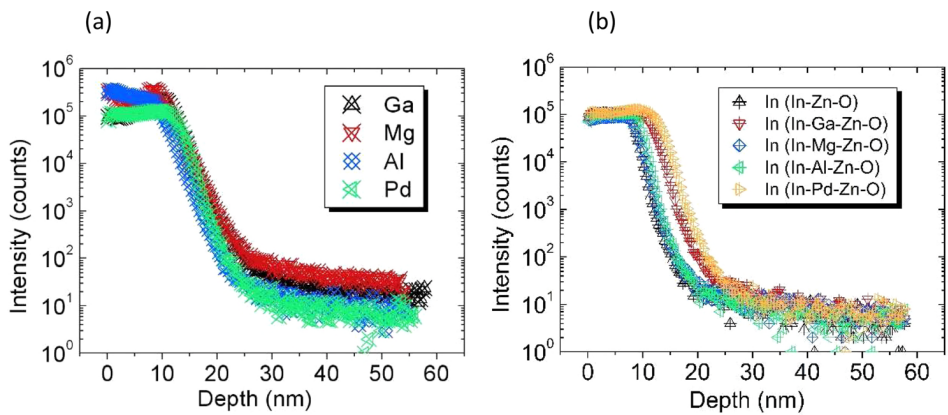

FIG. 3. Chemical and structural evolution of IXZO for (X = Ga, Mg, Al, Pd): (a) Profile of the dopants Ga, Mg, Al and Pd; (b) Profile of In in various IXZO films.

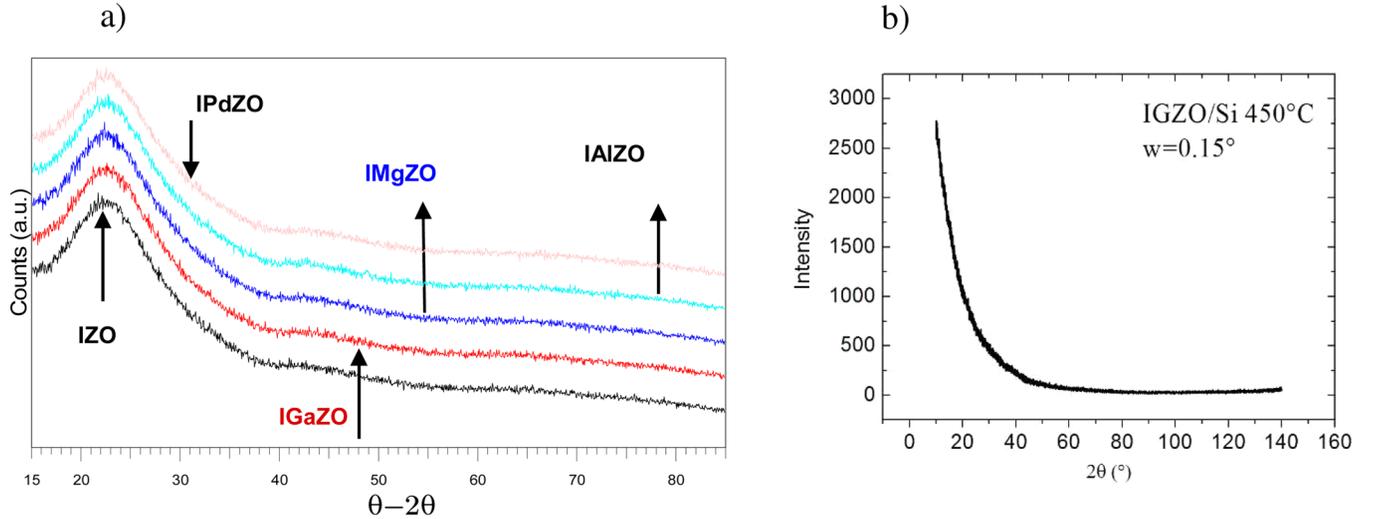

FIG. 4. (a) X-ray diffraction patterns of sol-gel In-X-Zn-O films on glass substrates annealed at 450 °C (b) X-ray patterns at grazing incidence angle ($\theta = 0.15°$) on silicon substrate.

## E. Optical UV-Vis Transmittance spectra analysis

IXZO films deposited on glass substrates were used for optical measurements. A baseline measurement, of the substrate transmittance was carried out using an ultraviolet–visible (UV-Vis) spectrophotometer Perkin Elmer LAMBDA 950. We analyzed the optical properties of the solution-processed IXZO films in the 250–1500 nm wavelength range. The prepared films have high average transmittance values in the visible light region, (over 90%), as shown in Figure 5(a).

The results shown in Figure 5(b) demonstrate that the doping at low concentrations does not affect significantly the energy band gap $E_g \approx 3.2$ eV, which is close to that of IZO. The value of $E_g$ was estimated by calculating the absorption coefficient using $\alpha = (1/t)\ln[1/T]$, where t is the film thickness (here 40 nm on glass substrate), and T is the measured transmittance. As known, the absorption coefficient of a direct-band gap semiconductor near the band edge is given by[23–25]

$$\alpha h\upsilon = D(h\upsilon - E_g)^{1/2}, \quad (2)$$

where $h\nu$ is the photon energy and D is a constant. The value of $E_g$ was obtained from the intercept of a linearly extrapolated $(\alpha h\nu)^2$ *versus* $h\nu$ plot to the energy axis, as shown in Figure 5(b) (dashed red line). A slight variation of $E_g$ with X doping is observed; $E_g$ varies from 3.15 eV (undoped In-Zn-O) to 3.3 eV for X = Ni.

## III. THIN FILM TRANSISTORS

### A. DC current-voltage characteristics

TFTs with 14 ternary IXZO compounds and one binary In-Zn-O were synthesized by the sol-gel process described above. Highly doped silicon substrates (p$^{++}$/Si) covered with thermal SiO$_2$ (100 nm thick) gate dielectric are used to fabricate TFTs with a bottom-gate bottom-contact architecture. Source and drain electrodes consisted of 30 nm Au layer deposited on top of 10 nm Ti anchor layer, and were patterned by optical lithography and lift-off technique. The TFT channel width to length ratio (W/L) is fixed at 500 (L = 20 $\mu$m). Current−voltage measurements were measured with Agilent model HP4156C semiconductor parameter analyzer using a probe station. All electrical characteristics were acquired in ambient conditions without intentional light blocking.

Electrical characterization of the TFTs consisted of taking drain-source current/gate voltage (transfer) characteristics $I_{DS}$ ($V_G$) at a constant $V_{DS}$, selected in the linear or the saturation region of the $I_{DS}(V_{DS})$ curve (Fig. 6). The $I_{DS}$ ($V_G$) characteristics were subsequently used for extracting the mobility, in low field (linear) region, $\mu_0$, and in the saturation regime, $\mu_{sat}$, respectively. The subthreshold slope, SS, and the threshold voltage, $V_{th}$, values, are known to be related to the trap density at the interface.[26] For the linear regime of the $I_{DS}(V_G)$ function, we have,

$$SS = \left(\frac{d\log(I_{DS})}{dV_G}\right). \quad (3)$$

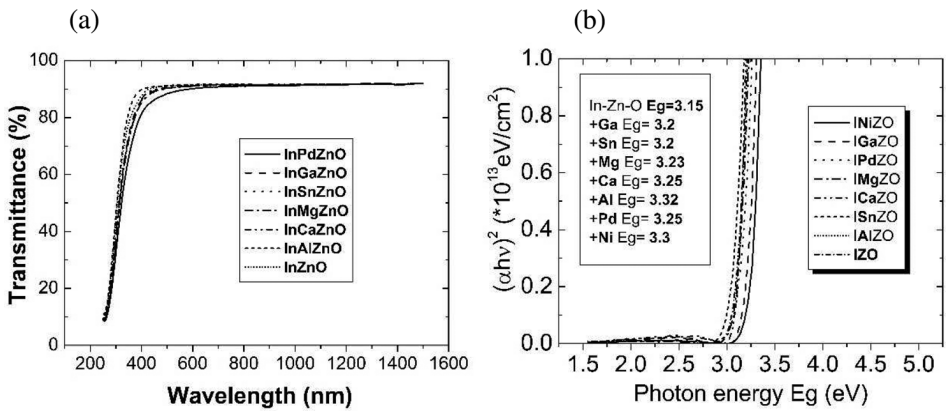

FIG. 5. (a) Transmittance spectra of the as-deposited amorphous IXZO with various dopants. (b) Extrapolated $E_g$ values for the same IXZO films in (a).

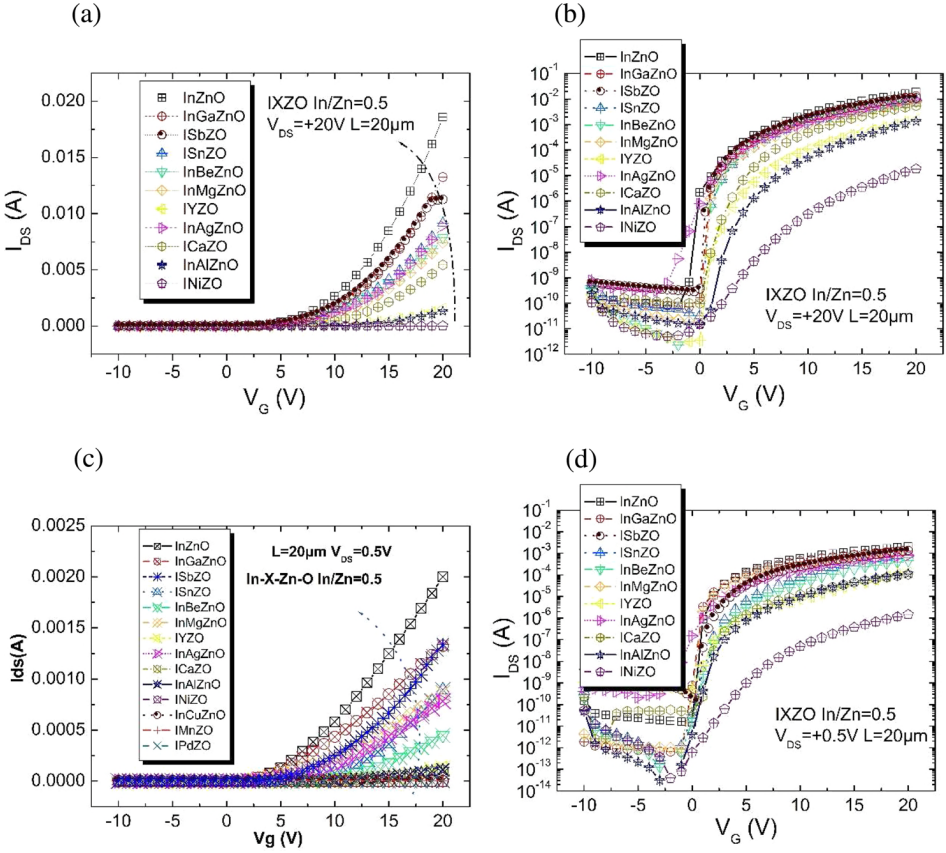

FIG. 6. Transfer characteristics of In-X-Zn-O devices in the saturation regime ($V_{DS} = 20$ V) (a) linear scale (b) log scale and linear regime ($V_{DS} = 0.5$ V) (c) linear scale (d) log scale.

From which the trap density can be found from the expression,

$$N_{ST} = \left(\frac{q}{(kT \times SS)} - 1\right)\frac{C_i}{q}, \quad (4)$$

where $C_i$ is the gate oxide capacitance, k the Boltzmann constant, T the temperature and q the elementary electron charge. Involved in the derivation of Eq. (4), it is the assumption that the interface trap density is uniform in energy and space.

In the saturation region the $I_{DS}(V_G)$ dependence takes the form,

$$I_{DS} = \frac{W}{2L}C_i\mu_{sat}(V_G - V_{th})^2, \quad (5)$$

where W and L are the channel width and length, respectively. The saturation effective mobility $\mu_{sat}$ was extracted from the slope of $\sqrt{I_{DS}}(V_G)$ plot at $V_{DS} = 20$ V, and the threshold voltage $V_{th}$ from the intercept of the extrapolated $\sqrt{I_{DS}}(V_G)$ plot with the horizontal, $V_G$ axis.

For the linear regime, at small $V_{DS}$, an alternative method of mobility extraction was used. As known,[26] the low field mobility $\mu_0$ can be evaluated from the slope of the function $Y(V_G) = I_{DS}/\sqrt{g_m}$, where $g_m = \partial I_{DS}/\partial V_G$ is the transconductance,

$$Y = \frac{I_{DS}}{\sqrt{g_m}} = \sqrt{\frac{W}{L}C_i\mu_0 V_{DS}}(V_G - V_{th}). \quad (6)$$

It should be noted that the method involving the use of the Y function yields the low field mobility $\mu_0$ and $V_{th}$ regardless of the access contact resistance.[27–29]

Table I summarizes the main TFT parameters obtained in the saturation regime for each IXZO TFTs. The mobility $\mu_{sat}$ and $V_{th}$ values were calculated using Eq. (5).

The $\mu$ and $V_{th}$ values obtained in the linear regime from the Y function are very similar to those presented in Table I (see Table II). The $\mu$ and $V_{th}$ values extracted from the data taken in the saturation and linear regimes are close, which proves not only that either region can be used for parameter extraction but also indicates a low contact resistance at source and drain. Device reproducibility was tested by measuring four identical transistors fabricated in four different batches. Device uniformity was found to be satisfactory (Figure 7), i.e., standard variations of $\mu$ and $V_{th}$ of less than 5% of the mean values and respectively.

Note that the dry–annealed (at 450 °C) IZO devices exhibit excellent properties, with $\mu$ of 8.5 cm$^2$V$^{-1}$s$^{-1}$ comparable with those reported in the literature for the Zn-rich compositions.[15]

We observe the following trends in the data presented in Table I (i) For X = Cu, Mn, Mo, and Pd the current intensity at $V_{DS} = 20$ V becomes very weak. These materials are highly resistive and useless for transistor fabrication; (ii) For IZO, IGZO, ISbZO, ISnZO, IBeZO, IMgZO, IAgZO, IYZO, ICZO, IAlZO, and INiZO, we observe a steady decrease in the mobility from 8.5 cm$^2$V$^{-1}$s$^{-1}$ (Ga) down to 0.8 cm$^2$V$^{-1}$s$^{-1}$ (Ni) with the dopant introduction as well as an associated increase in the threshold voltage. (iii) Generally, the $I_{off}$ values remain low for all the devices, at a value of about $10^{-10}$A, with the exception of Cu, $I_{off} \approx 10^{-9}$A.

TABLE I. Main electrical parameters of IXZO TFTs at saturation ($V_{DS} = 20$ V).

| In/Zn = 0.5 | $\mu_{sat}$(cm$^2$V$^{-1}$s$^{-1}$) | $V_{th}$(V) | $I_{on}$(mA) | $I_{off}$(A) |
|---|---|---|---|---|
| In-Zn-O | 8.5 | 3.2 | 19 | $3 \times 10^{-10}$ |
| In-**Ga**-Zn-O | 8 | 3.5 | 14 | $2 \times 10^{-10}$ |
| In-**Sb**-Zn-O | 7.7 | 3.5 | 12 | $2 \times 10^{-10}$ |
| In-**Sn**-Zn-O | 7 | 3.6 | 9 | $6 \times 10^{-10}$ |
| In-**Be**-Zn-O | 6 | 4.0 | 9.5 | $5 \times 10^{-10}$ |
| In-**Mg**-Zn-O | 5.5 | 4.6 | 8 | $4 \times 10^{-10}$ |
| In-**Ag**-Zn-O | 5 | 5.5 | 7.8 | $10^{-10}$ |
| In-**Y**-Zn-O | 4 | 6.5 | 7.4 | $8 \times 10^{-10}$ |
| In-**Ca**-Zn-O | 3 | 9.5 | 5 | $2 \times 10^{-10}$ |
| In-**Al**-Zn-O | 1.8 | 10 | 1.5 | $2 \times 10^{-10}$ |
| In-**Ni**-Zn-O | 0.8 | 12 | 0.1 | $10^{-10}$ |
| In-**Cu**-Zn-O | $10^{-4}$ | >20 | $2 \times 10^{-3}$ | $2 \times 10^{-9}$ |
| In-**Mn**-Zn-O | — | — | — | $<10^{-10}$ |
| In-**Pd**-Zn-O | — | — | — | $<10^{-10}$ |
| In-**Mo**-Zn-O | — | — | — | $<10^{-10}$ |

### B. Interface state density extraction

It should be emphasized that the transistor characteristics involve electron transport in the vicinity of the IXZO channel/gate oxide interface SiO$_2$. That is not only a consequence of the MOSTET configuration of the studied structures, but also of their "normally off" configuration. For this reason, we do not discuss conductivity of the channel materials, but the mobility of the current carriers in the channel alone. As the carriers remain close to the semiconductor/oxide interface, the feature of interest is the defect density in its immediate neighborhood. We evaluated the latter by two independent techniques, one using static characteristics of the TFTs (see above) and another using data extracted from the low frequency noise (LFN) measurements.

#### 1. Low frequency noise

Low frequency fluctuations in drain current in classical MOSFETs are usually interpreted by charge mobility fluctuations ($\Delta\mu$ model) and fluctuations in their density (carrier number fluctuation model, $\Delta$n). The latter is predominant in most field effect devices and originates from trapping-release of charges on the oxide traps localized near the channel semiconductor/gate-oxide interface. The charge trapping/detrapping causes flat-band voltage fluctuations which affect the charge carrier density in the channel ($\Delta$n model). The charge exchange between the traps and the channel proceeds by tunneling with the characteristic time constant, $\tau = \tau_0 \exp(-x/\lambda)$, where $\lambda$ is the tunneling constant (about 1 Å for SiO$_2$), and x is the trap-interface distance. For a uniform distribution of traps in the oxide, the power spectral density (PSD) of drain current fluctuations, $S_{ID}(f)$, falls as 1/f, where f is the frequency. Such dependence is observed in a vast family of semiconductor devices and can be used for evaluating the surface density of traps active in the tunneling process, which, in the first approximation, reside in the SiO$_2$ layer whose thickness is $\lambda$.

In the following, we disregard trapping processes in the semiconductor bulk, as, first, the current flow is confined to the near-interface region and, secondly, such effects occur at frequencies higher than the low frequency spectral window we used for LFN measurements. We also assume that mobility fluctuations do not contribute significantly to the current fluctuations. That effect can be disregarded when defect scattering is predominant over lattice (phonon) scattering, which can be expected in amorphous structures. Although transport processes in the oxide semiconductors are little known, such assumptions are justified.[30] For all these reason, we interpret our noise data using expressions appropriate for the $\Delta$n noise model mechanism.

We have carried out the low frequency noise, LFN, measurements in the IXZO channel devices described in the former sections, for 1 Hz < f < 10 kHz. Such frequency range is adequate for IXZO devices because their possible applications are also limited to low frequencies, owing to the low mobility of the channel carriers in the IXZO materials. We found that PSD for all the devices varied as 1/f, at sufficiently high $I_{DS}$ intensities and followed the $I_{DS}^\alpha$ dependence with $\alpha \approx 2$. An example of such data is shown in Figures 8(a) and 8(b).

TABLE II. Summary of electrical characteristics of the IXZO channel transistors extracted from the LFN data at f = 16 Hz and from the static (linear regime $V_{DS} = 0.5$ V) data for various dopants.

| IXZO for In/Zn = 0.5 | SS$^{-1}$ (V/decade) | $V_{th}$ (V) | $N_{ST}$(LFN) (10$^{12}$eV$^{-1}$cm$^{-2}$) | $N_{ST}$ (Stat-Lin) (10$^{12}$eV$^{-1}$cm$^{-2}$) | $\mu_0$ (cm$^2$V$^{-1}$s$^{-1}$) |
|---|---|---|---|---|---|
| In-Zn-O | 0.22 | 3.2 | 1.4 | 1.7 | 9 |
| In-**Ga**-Zn-O | 0.23 | 3.5 | 1.4 | 1.8 | 8.5 |
| In-**Sb**-Zn-O | 0.24 | 3.5 | n.m | 1.7 | 8 |
| In-**Sn**-Zn-O | 0.25 | 4 | n.m | 2.4 | 7 |
| In-**Be**-Zn-O | 0.24 | 3.6 | n.m | 2.4 | 6.5 |
| In-**Mg**-Zn-O | 0.23 | 4.6 | 2.4 | 3.4 | 6 |
| In-**Ag**-Zn-O | 0.50 | 5.5 | n.m | 4.2 | 5.7 |
| In-**Y**-Zn-O | 0.85 | 6.5 | n.m | 4.5 | 4 |
| In-**Ca**-Zn-O | 0.9 | 9.5 | 9 | 5.4 | 2 |
| In-**Al**-Zn-O | 0.9 | 10 | 31 | 7 | 1.5 |
| In-**Ni**-Zn-O | 2.5 | 12 | 51 | 26 | 0.9 |
| In-**Cu**-Zn-O | 9.0 | >20 | — | 75 | $3 \times 10^{-4}$ |

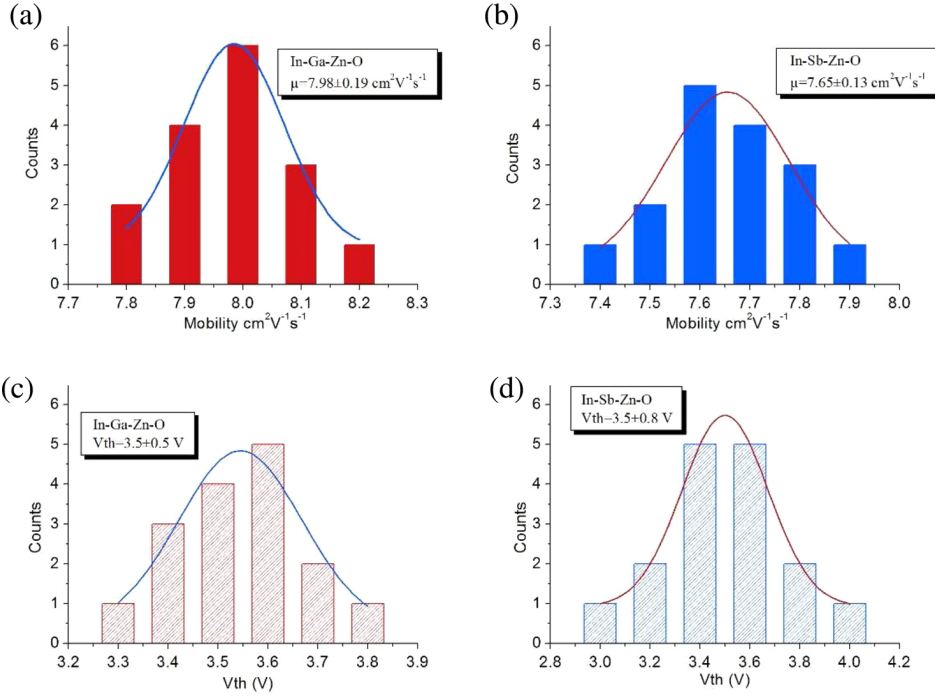

FIG. 7. Histograms of (a) mobility and (b) Vth for IGZO (c) mobility and (d) Vth for ISbZO in saturation regime. The mean field effect mobility ($\mu$) in saturation region (at $V_{DS} = +20$ V) is 8 cm$^2$ (V.s)$^{-1}$ and Vth = 3.5 V for In-Ga-Zn-O and is $\mu = 7.7$ cm$^2$(V.s)$^{-1}$ and Vth = 3.5 V for In-Sb-Zn-O. The standard deviations are 0.19 and 0.13 cm$^2$V$^{-1}$s$^{-1}$, i.e., a relative dispersion less than 5% of the mean value.

The quantity of interest for extracting the density of noise sources, is $S_{ID}(f)/I_{DS}^2$ with $S_{ID}$ being measured at a given frequency, $f_0$, where the 1/f law is valid (here $f_0 = 16$ Hz). The variation of $S_{ID}(f)$ as 1/f and $S_{ID}(I_{DS})$ a s $I_{DS}^2$ justifies the use of the $\Delta$n-model for data interpretation. In this case,[31,32]

$$\frac{S_{ID}}{I_{DS}^2} = S_{VFB} \times \left(\frac{g_m}{I_{DS}}\right)^2 = \frac{q^2 kT N_{ST}}{WLC_i^2 f} \times \left(\frac{g_m}{I_{DS}}\right)^2, \quad (7)$$

where $g_m = \partial I_{DS}/\partial V_G$ is the transconductance. The constant multiplying $(g_m/I_{DS})^2$ in Eq. (7), corresponds to the flat band voltage PSD and contains the trap density value, $N_{ST}$. The latter can be directly obtained by superimposing the plots of the functions $S_{ID}/I_{DS}^2$ *versus* ($I_{DS}$) and constant $\times (g_m/I_{DS}^2)$ *versus* ($I_{DS}$), until the graphs merge, adjusting the constant as illustrated in Figure 8(b). The feasibility of such a proce-dure also confirms the validity of the $\Delta$n model of noise generation. The LFN data obtained on the IXZO TFT devices showed a predominance of the carrier-number fluctuation noise generation, which justified the universal use of the $N_{ST}$ extraction method described above. It is very likely that this method could be generally applied to In/Zn oxide devices. We discuss the LFN data together with the static data in the following section as both show the same behavior.

### 2. Interface trap density from TFT static characteristics and LFN

In Figure 9, the $N_{ST}$ values obtained from the static IV characteristic (Eqs. (3) and (4)) and LFN data are plotted together for various X dopants. The series of the data points was arranged in an increasing linear sequence of $N_{ST}$. It can be readily seen that the LFN and static measurements yield similar $N_{ST}$ values and show the same trend in the data arrangement in Figure 9. The lowest $N_{ST}$, of around $10^{12}$ eV$^{-1}$cm$^{-2}$ was obtained for IZO and IXZO with X = Ga and Mg. the highest trap density, above $10^{13}$ eV$^{-1}$cm$^{-2}$ was found for devices with X = Ca, Al, Ni.

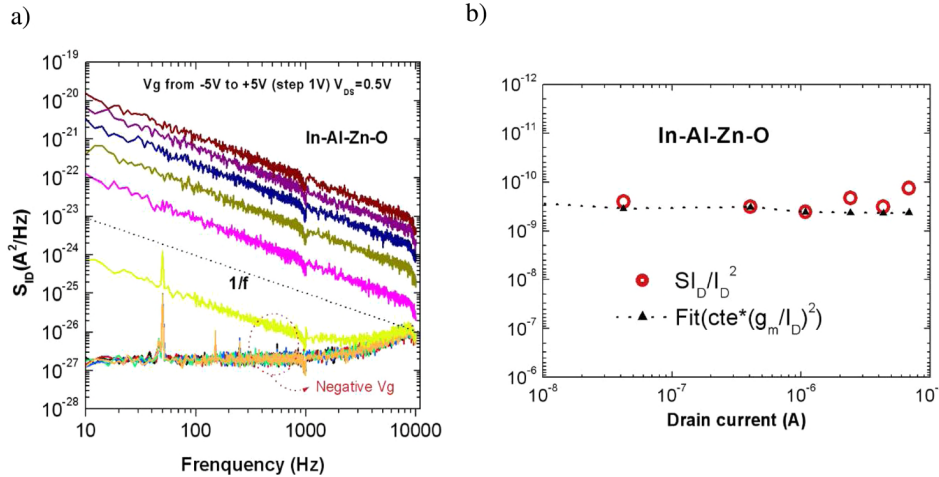

FIG. 8. (a) The measured PSD of LFN in a InAlZn0 in the linear regime, the dash line indicates the 1/f slope. The increase in $S_{ID}$ at high frequencies and at small gate voltages is due to the instrument limitations. The PSD values at 16 Hz are used for the subsequent analysis. (b) Superimposition of $S_{ID}/I_{DS}^2$ and const. $\times (g_m/I_{DS}^2)$ plots. Note that $S_{ID}/I_{DS}^2$ is practically constant, which follows from the observation that $S_{ID} \sim I^2$ (not shown).

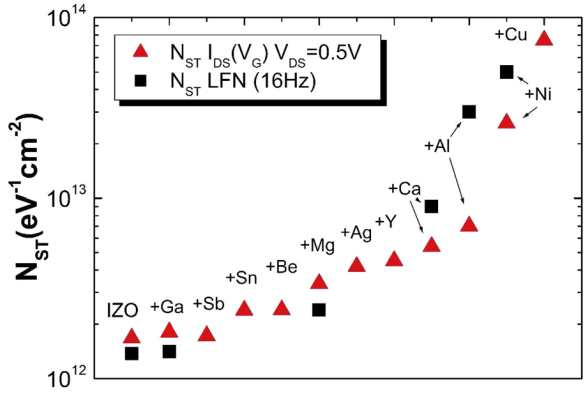

FIG. 9. Interface trap density, $N_{ST}$ determined from SS in the linear region of $I_{DS}(V_G)$, for $V_{DS} = 0.5$ V (triangles) and from the PSD of LFN at f = 16 Hz (squares).

Table II summarizes some characteristics of the studied IXZO channel TFTs.

### 3. Correlation between $N_{ST}$, $V_{th}$, and $\mu$ values of the IXZO channel TFTs

The $N_{ST}$ data presented in Figure 10 are shown together with the corresponding low field mobility data (respectively red and black data points on line) in order to demonstrate clearly how drastically the carrier transport in the IXZO is affected by traps. The mobility decrease with the growing defect density, such as shown in Figure 10, is consistent with the data reported in the literature on IZO and IGZO TFTs.

In particular, Lee et al.[38] have applied the trap-limited conduction and percolation conduction models to interpret their transport data in the structures involving IGZO/IZO layers. The origin and the localization of traps in IGZO fabricated by vacuum techniques have also been discussed in a paper by Kamaya et al.[39] in the context of trap-limited electron transport. The mobility decrease with increasing trap density is ubiquitous in semiconductor materials and was studied in detail in classical semiconductors at high dopant levels and at high compensations, where the mobility degradation can severely limit device performances.

Such behavior is also typical for amorphous materials, such as α-Si, thus some basic concepts developed in the studies of the latter apply to the IXZO materials. We shall refer to the band structure of amorphous materials in the discussion of the IXZO properties that we address presently.[33–37] In the TFTs with amorphous IXZO channels, we are concerned with the structure of the conduction band at the SiO$_2$/IXZO interface. Potential fluctuations, due to imperfections at the latter, induce localization of the states near the conduction band minimum at the oxide side, leading to the appearance of a band tail below its bottom. The tail states are localized (transport by hops) up to certain energy $E_c$ (mobility edge) and are delocalized above it. In other words, at $E > E_c$ transport is carried by free electrons, with a well-defined mobility $\mu_f$. The total number of electrons in the system is thus divided into the localized and free states. Therefore, the measured mobility, $\mu_0$, in the system is lower than the mobility of the free carriers, $\mu_f$; it is scaled by a factor comprising the free and localized electron concentrations, respectively, $n_f$ and $n_l$, thus

$$\mu_0 = \mu_f \times \frac{n_f}{n_f + n_l}. \quad (8)$$

Such mobility degradation with the trap density increase can be clearly visualized by plotting $\mu$ versus $N_{ST}$, as shown in Figure 11. The solid line corresponds to an empirical exponential fit of the form:

$$\mu_{fit} = \mu_0 \times \exp\left(-\frac{N_{ST}}{N_{TC}}\right), \quad (9)$$

where $N_{TC}$ is a trap density parameter (here $N_{TC} \approx 4 \times 10^{12}$/eVcm$^2$ from fitting data in Fig. 11). Its value is consistent with previously reported data for a-IZO/IGZO TFTs.[38]

Similarly, as it can be seen from Figure 12, the variation in the charge trapped at the interface also explains the threshold voltage shift with $N_{ST}$. As known, the threshold voltage in a MOSFET can be written as,

$$V_{th} = V_{th0} + \frac{Q_{trap}}{C_i}, \quad (10)$$

where $Q_{trap} = q.N_{ST}.\Delta\Psi_s$, with $\Delta\Psi_s$ being the surface potential sweep from weak to strong inversion. The calculation with $\Delta\Psi_s = 0.25$ eV accounts satisfactorily for the data.

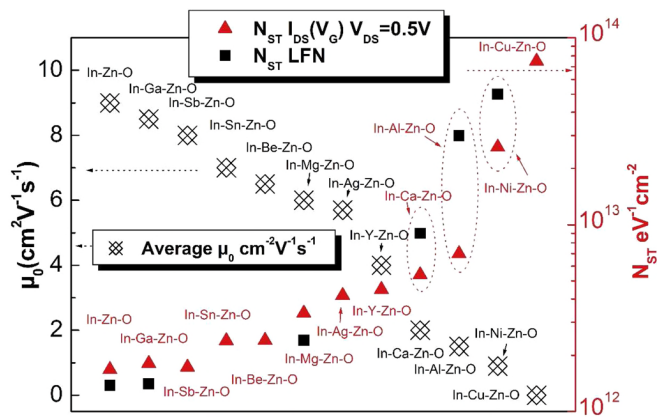

FIG. 10. Compilation of mobility $\mu$ (black diamonds)) and surface trap density data $N_{ST}$ (black squares) and SS data (red triangles) in transistors with IXZO channels. An increase in $N_{ST}$ is associated with a decrease in $\mu$.

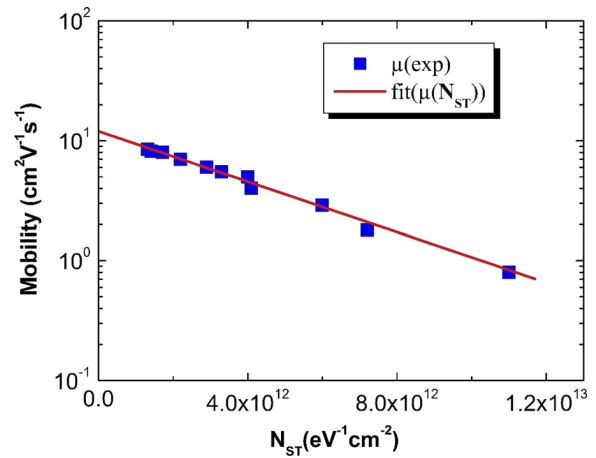

FIG. 11. Variation of the low field mobility with the interface trap density $N_{ST}$ in transistors with IXZO channels. Exponential fit is obtained from Eqs. (8) and (9).

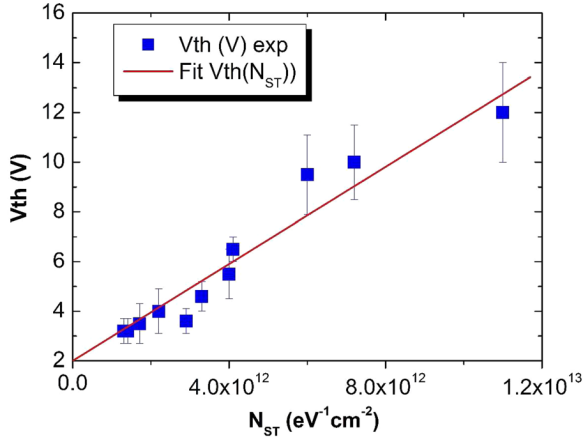

FIG. 12. Variation of the threshold voltage with the trap density obtained for several transistors with IXZO channels. The solid line results from a calculation involving Eq. (10), with $\Delta\Psi_s = 0.25$ eV.

An exponential decrease of mobility with trapped interface charge is consistent with an increase of potential fluctuations-induced disorder as in amorphous semiconductors.[33–35,38]

## IV. DISCUSSION

We carried out a systematic study of properties of IXZO ternary oxides with dopants chosen to have an appreciable dispersion in the ionic radii values and in their electronegativity. The layers were first analyzed using several standard material characterization techniques, such as XPS, DRX, SIMS, and their transmission spectra in the UV-visible light were determined. All the layers had amorphous structure, were uniform and smooth. The direct gap value of the IXZOs was found to be independent of the dopant species. In view of the absence of changes in IXZO's structural and morphological properties with X, the gap variations were disregarded in our interpretation of electrical properties of the IXZO TFTs. In order to understand why the electrical performances of the TFTs with IXZO channel materials depend on the dopant species, we addressed the question of chemical and morphological structures of the IXZO ternaries in conjunction with the defect generation in the structures and charge carrier mobility of the TFTs fabricated with IXZO channels.

The mobility degradation in the TFTs was observed to be particularly dramatic for X = Cu, Mn, Pd, and Mo. It is reasonable therefore to assume that these dopants induce morphological changes in the IXZO structures. However, our DRX studies have not revealed any significant changes of this type upon doping and the structures remained amorphous. Moreover, the XPS analysis revealed that all the oxides had a similar composition, described by the generic formula, $In_{0.5}X_{0.08}ZnO_{2.5}$ (about 10 at. % of In, 20 at. % of Zn, and 1.6 at. % of X). Another factor detrimental for mobility is the presence M-OH bonds in the structure.[20] The latter was observed at small, trace quantities in all the XPS spectra, regard-less of the dopant, exact evaluation of the concentrations of M-OH traps, let alone their variation from sample to sample, was difficult because they rested in the limit of detection of the technique used.

Hennek et al.[40] have presented results on the effects of doping of IXZO by Ga, Sc, Y, and La ions, having strong oxygen affinity ("gettering capacity") and various ionic radii values, namely $IR_{Ga} = 0.76$ Å, $IR_{Sc} = 0.89$ Å, $IR_Y = 1.04$ Å, and $IR_{La} = 1.17$ Å, respectively. One of the objectives of Hennek et al.[40] was to test whether the oxygen vacancy creation (oxygen gettering) or the structure amorphization, following the doping, were responsible for the observed mobility decrease. They concluded that the mobility drop was generally higher in IXZO with larger $IR_X$ and always decreased when the dopant concentration exceeded about 2.5 at. %. A part of our study is an extension of the work of Ref. 40 over a wider range of dopants in search of correla-tions between the electrical properties of the studied ternary IXZO's and the $IR_X$ values of the dopants. We found a clear correlation between $\mu$ and IR values only for the alkaline earths metals, Be, Mg, Ca, whose dopant radii were, respec-tively, $IR_{Be} = 0.45$ Å, $IR_{Mg} = 0.65$, and $IR_{Ca} = 0.95$. Namely, we found that the mobility values in the Be, Mg, and Ca doped IXZO decreased with the increasing IR of the dopant ions. However, no correlation between the values of IR and the values of $\mu$ could be revealed in the TFTs with other ions. For example, comparing TFTs with IGZO and IAlZO chan-nels for the same (low) dopant concentrations, fabricated using the same technology, we found that the TFTs with Ga-doped channel showed mobility values of 8 cm$^2$V$^{-1}$s$^{-1}$ while in Al-doped TFTs $\mu$ did not exceed 1.8 cm$^2$V$^{-1}$s$^{-1}$, notwithstanding the closeness of their ionic radii ($IR_{Ga} = 0.6$ and $IR_{Al} = 0.5$ Å).[17] On the other hand, our mo-bility data (Table I) also showed that similar $\mu$ values were obtained for IXZO, at low X concentrations, for X whose ionic radii differed significantly. For example, for IBeZO ($IR_{Be} = 0.4$ Å) $\mu$ values of 6 cm$^2$V$^{-1}$s$^{-1}$ was obtained, while for the large IR elements such as Sn ($IR_{Sn} = 0.7$ Å) or Sb ($IR_{Sb} = 0.9$ Å) the $\mu$ values were, respectively, 7 cm$^2$V$^{-1}$s$^{-1}$ and 7.8 cm$^2$V$^{-1}$s$^{-1}$, thus comparable. The search for correla-tions between the ionic radii of the dopants and the TFT mo-bility has not led us to disclosing any simple general rule applicable to all the dopants used. As discussed, some trends were observed for a limited number of X atoms, sharing cer-tain common characteristics.

We also addressed a question of how the electronegativity of the dopant may affect the properties of the synthesized IXZO. It should be reminded here that the difference in the electronegativity of the dopant and the matrix metals, Zn and In, should play an important role at the early stage of the sol-gel synthesis.[41] Namely, atoms of lower electronegativity bind more readily with to O atoms. We could identify three groups of dopants, according to the mobility values in the TFTs with IXZO channels,

(1) The highest $\mu$ values were found for Ga, Sb, and Sn dopants. Their electronegativity is the highest and superior of that of In and Zn.
(2) The atoms giving the intermediate $\mu$ values (Be, Mg, Ag, Y, Ca) had the electronegativity below that of In.
(3) The lowest $\mu$ values were found for metallic X, namely Ni, Cu, Mn, Pd, and Mo.

The properties of the first group of ions may be explained by observing that the dopants with the electronegativity higher than that of Zn and In perturb less the In-O and Zn-O sequences during the IXZO formation in the liquid phase. Ions in the second group, with lower electronegativity, interfere into the hydrolysis and other condensation reactions, which disturbs more the formation of the In-O and Zn-O sequences.[42] As to the third group of dopant atoms, they most probably enter in the ternary system during the synthesis already as clusters or inclusions in the resulting solid, forming defects.

Using standard analysis of the $I_{DS}(V_G)$ characteristic of the TFTs, we obtained carrier mobility values in the transistor channels, together with the density of the electrically active defects, located near the channel semiconductor/gate oxide interface. The values of mobility obtained from the saturation and linear regions of transistor operation were close, hence only the low-field mobilities were further analyzed. The interface defects density, obtained from two independent measurement techniques, (i) conventional technique, involving $I_{DS}(V_G)$ subthreshold slope and $V_{th}$, and (ii) involving analysis of the LFN data, coincided within the experimental error bar. That confirms the localization of the electrically active defects at the SiO$_2$/IXZO interface. An arrangement the $\mu$ values for all the dopants in the descending order revealed that the corresponding trap densities, $N_{ST}$, followed an increasing trend and the values of $\mu$ and $N_{ST}$ are linked by an exponential relation (cf. Eq. (9)). That is attributed to the formation of a band of localized states near the conduction band minimum whose width is proportional to the localized charge density, as in dis-ordered semiconductor materials.[33,35] Thus, we can conclude that the main factor governing the charge carrier mobilities in the IXZO TFTs is the interface defect density. The lowest interface defect density (therefore the highest TFT mobilities) was obtained for X dopants with the highest electronegativity and ionic radii close to that of In and Zn.

## V. CONCLUSIONS

This systematic study on the role of the dopants X in the sol-gel fabrication of the IXZO ternaries stemmed from the need of assessing what dopant features affect most the synthesis processes and the properties of its products. For this reason, we used a significant number of atomic species, selecting them to have a wide spectrum of electronegativity and IR values, as those are known to be essential in the sol-gel technology. We found that the ions with electronegativity slightly superior to that of In and Zn (Ga, Sn, and Sb) give the lowest $N_{ST}$ values and consequently the best field-effect $\mu$ in the TFTs that were also fabricated in the course of this work. The IR length seems to play a secondary role in the defects creation, however, the ions with IR close to those of In and Zn should stress the resulting solid less, creating less defects. The best $\mu$ and the lowest $N_{ST}$ were found in Ga, Sb, and Sn-doped In/Zn oxide. They should be recommended to use in the sol-gel technology of the oxide-based transistors and displays. Our results showed an exponential drop of $\mu$ as a function of $N_{ST}$. This is in agreement with the existence of a band of localized states near the conduction band minimum of the oxide, as in the classical disordered semiconductors.

Finally, it should be pointed out that electrical characteristics of the TFT fabricated with Ga, Sn and Sb-doped IXZO are quite satisfactory and showed no drift in storage or during the test measurements. The highest mobility, of around $8\,cm^2 V^{-1} s^{-1}$ and the lowest $N_{ST}$, around $10^{12}\,eV^{-1} cm^{-2}$ was found for IGaZO, but Sb and Sn-doped transistors are nearly as good. From a technological perspective, our work points out on the dopants best adapted for stabilizing the IXZO structure. The Sn and Sb dopants give devices with characteristics similar to those of IGZO. They may become the winning contenders for Ga-doped IXZOs for applications envisaged for the transparent semiconductor oxide devices.

## ACKNOWLEDGMENTS

The authors are grateful to Christophe Serbutoviez and to Olivier Poncelet and to Jean-Emmanuel Broquin for discussions on sol gel and glass structures.